\documentclass[useAMS,usenatbib]{mn2e}

\usepackage{graphicx}
\usepackage{amssymb}
\usepackage{color}
\usepackage{pst-grad}
\usepackage{psfrag}

\newcommand{\pdiff}[2]{\frac{\partial #1}{\partial #2}}

\newcommand{\eqb}{\begin{eqnarray}}
\newcommand{\eqe}{\end{eqnarray}}
\newcommand{\diff}{{\rm d}}
\newcommand{\bfm}[1]{\mbox{\boldmath$ #1 $}}
\newcommand{\QLT}{quasi-linear theory} %Stop changing this. It puts a space before , and .

\newcommand{\Add}{\textbf{W} }
\newcommand{\Bdd}{\textbf{X} }
\newcommand{\Cdd}{\textbf{Y} }
\newcommand{\Ddd}{\textbf{Z} }

\newcommand{\Aod}{\textbf{W1} }
\newcommand{\Bod}{\textbf{X1} }
\newcommand{\Cod}{\textbf{Y1} }
\newcommand{\Dod}{\textbf{Z1} }

\newcommand{\Atd}{\textbf{W3} }
\newcommand{\Btd}{\textbf{X3} }
\newcommand{\Ctd}{\textbf{Y3} }
\newcommand{\Dtd}{\textbf{Z3} }

\newcommand{\n}{\noindent}
\newcommand{\beq}{\begin{equation}}
\newcommand{\eeq}{\end{equation}}

\newcommand{\gesim}{\,\raisebox{-0.4ex}{$\stackrel{>}{\scriptstyle\sim}$}\,}
\newcommand{\lesim}{\,\raisebox{-0.4ex}{$\stackrel{<}{\scriptstyle\sim}$}\,}

\newcommand{\mb}[1]{\bfm{#1}}

\newcommand{\rg}{{r}_{\rm g}}
\newcommand{\og}{{\omega}_{\rm g}}
\newcommand{\ogi}{\og^{-1}}

\title[Stochastic particle acceleration in radio lobes]
{Stochastic particle acceleration in the lobes of giant radio galaxies}
\author[O'Sullivan, Reville, Taylor]{ 
S.~O'Sullivan$^1$, B.~Reville$^2$, A.~M.~Taylor$^2$\\
$^1$School of Mathematical Sciences, Dublin City University, Dublin, Ireland\\
$^2$Max-Planck-Institut f\"ur Kernphysik, 69029, Heidelberg, Germany}
\begin{document}

\date{Accepted 2009 July 22. Received 2009 July 16; in original form 2009 March 05}

\pagerange{\pageref{firstpage}--\pageref{lastpage}} \pubyear{2009}

\maketitle

\label{firstpage}

\begin{abstract}
We investigate the acceleration of particles by Alfv\'{en} waves via the second-order Fermi process
in the lobes of giant radio galaxies. Such sites are candidates for the 
accelerators of ultra-high energy cosmic rays (UHECR). We focus on the nearby FR~I
radio galaxy Centaurus~A. This is motivated by the coincidence of its position with the arrival direction of several of the highest energy Auger events.
The conditions necessary for consistency with the acceleration timescales predicted by quasi-linear theory are reviewed. Test particle calculations
are performed in fields which guarantee electric fields with no component parallel to the \emph{local} magnetic field. The results of \QLT~are, to order of magnitude, found to be accurate at low turbulence levels for non-relativistic Alfv\'{e}n waves and at both low and high turbulence levels in the mildly relativistic case. 
We conclude that for pure stochastic acceleration via Alfv\'{e}n waves to be plausible as the generator of 
UHECR in Cen~A, the baryon number density would need to be several orders of magnitude below currently held upper-limits.
\end{abstract}

\section{Introduction}

The origin of ultra-high energy ($\gesim 10^{18}$~eV) cosmic rays (UHECR) remains a long standing mystery
in high energy astrophysics. Few objects are plausible as potential sources. In the Milky Way's $\sim\mu$G magnetic fields, for example, UHECR have gyroradii comparable to, 
or larger than, the size of the Galaxy. Moreover, the apparent isotropy in their arrival
directions, at least
up to 10$^{19.7}$~eV, indicates that they must have propagated a sufficient distance
to have undergone multiple scatterings. This clearly
points to an extra-galactic origin of UHECR.

Consideration of the possible sources for these particles has resulted in size, kinetic energy, and magnetic field strength constraints \citep{Hillas:1984}, 
with it being possible to infer little more beyond this.
Recent observations by the Pierre Auger Observatory (Auger) have indicated
that UHECR with energies $>10^{19.7}$~eV, appear to have statistically significant
(at the 3$\sigma$ level) correlations with the local ($<75$~Mpc) AGN distribution in the Veron-Cetty catalogue AGN population 
\citep{Abraham:2008}. No such correlation was found in HiRes events 
\citep{HiRes}, although the location of the observatory in the northern hemisphere required
analysis with a different AGN set from the catalogue. 
Nevertheless, such a signal is particularly interesting since it resides
in the energy range where the propagation of UHECR becomes limited by GZK interactions 
\citep{Greisen:1966jv,Zatsepin:1966jv}, to several hundreds of Mpc in distance; 
an effect which should make any such correlation easier to see.

One of the properties of cosmic rays at ultra-high energies
is that, for $<$nG extragalactic fields, they are capable of propagating great distances (tens to hundreds of Mpc) in the intergalactic medium almost rectilinearly.
One source of particular interest is Centaurus~A, a FR~I radio galaxy 
in the southern hemisphere which has been associated with a number of the
Auger detections above $10^{19.7}$~eV: two with the centre itself
\citep{Abraham:2008} and additional others with the extended lobes~\citep{Mosk08}.
However, whether Cen~A is merely deflecting particles towards 
our line of sight, or is itself the site of acceleration is a topic
of current interest~\citep[e.g.][]{Lemoine, Gorb}.

Within the AGN class of objects, several different acceleration sites present
themselves as candidate locations for the acceleration of UHECR. Among these are 
the central AGN, the jet driving the lobe, 
and the large radio lobes themselves~\citep[for a review see][]{BBR84}. 
Furthermore, several acceleration mechanisms may be acting
in parallel, with different processes dominating depending on the local conditions~\citep{Rieger:2006}.
In particular, while first-order Fermi processes may be occurring near shocks, shear or
inductive acceleration can operate in the jet~\citep{Ostro98,Rieger:2004,Lyutikov}, and stochastic acceleration is likely to occur within the lobes.

In this paper we consider the case for stochastic acceleration being responsible for the production of UHECR in giant radio galaxies. The lobes of these radio galaxies are
most likely highly turbulent environments, where the fractional energy density
in the magnetic turbulence, $\delta B^{2}/ B_0^{2}$, is not necessarily a small parameter. 
Due to its size and estimated field strength, 
it has been argued by \citet{Hardcastle} that stochastic acceleration may  be a possible mechanism for accelerating UHECR above $10^{19}$~eV in Cen~A. It was assumed in this work that the magnetic energy density, $B_{tot}^2/8\pi$, was within
$10\%$ of the rest mass energy density of the thermal
protons in the lobe $0.5n_{\rm p} m_p c^2$. 

Earlier numerical studies of stochastic acceleration were 
mostly concerned with methods for solving the particle transport equation
\citep[e.g][and references therein]{Park96}, or the use of stochastic differential equations
\citep{Acht92}.
Since tracking large ensembles of test particles is 
computationally expensive, previous investigations considered either the short time dependencies of diffusion characteristics on turbulence level in a narrow dynamic range
of $k_{\rm max}/k_{\rm min}=10^{3}$~\citep{MichOst96, MichOstSchlick99}, or rapid acceleration in fully turbulent strong fields \citep[e.g][]{Arzner}. Attention has also been directed towards the influence of coherent acceleration processes invoked by means of an electric field component parallel to the local magnetic field~\citep{Dmitruk, Arzner}. While parallel fields may be legitimately admitted via resistive processes in the MHD fluid approximation, \cite{Arzner} have pointed out that if an electric field is generated via the Lorentz transformation of time dependent Fourier modes representing magnetic fluctuations, an underestimate of the acceleration times may result as a consequence of finite local parallel electric fields.

An attempt at following the turbulent acceleration of protons in radio lobes 
over many decades in energy has recently been presented
in \citet{FF08}, although the acceleration timescales resulting from their numerical scheme
appear to disagree with quasilinear theory predictions by several orders of magnitude. 
Furthermore, their results, taken at face value, seem to be in conflict with
the Hillas criterion, which limits the maximum particle energy $E_{\rm max}$ 
to which a given source can accelerate, $E_{\rm max}=Ze \beta_{\rm A} Bc R$, where 
$Z$ is the charge number, $\beta_{\rm A}$ is the Alfv\'en velocity
in units of $c$, $B$ is the magnetic field strength, and $R$ is the
size of the source.

The goal of this paper is to use quasilinear theory to place constraints
on the parameters required to accelerate protons or ions to $10^{18}$~eV
and above, assuming
a homogeneous turbulent radio lobe. Numerical simulations are performed to
check the accuracy of quasi-linear theory results when $\delta B^{2}/B_0^{2}$ 
and $\beta_{\rm A}$ are not small parameters.
The outline of the paper is as follows.
In Section~\ref{acceleration} we review the basic results of
stochastic acceleration using quasilinear theory, and discuss 
limitations on the acceleration timescales for UHECR. Section~\ref{CenA} 
focuses directly on the acceleration of protons and 
electrons in the giant outer radio lobes of Cen~A. In Section~\ref{numerics} we discuss the numerical model, demonstrating to what degree
the results of quasilinear theory hold in highly turbulent regions with non-linear magnetic turbulence or mildly relativistic Alfv\'en velocities. 
Our conclusions are presented in Section~\ref{conclusion}.

\section{Particle acceleration in turbulent fields}
\label{acceleration}

The motion of a charged particle, with charge $q$, in electric 
and magnetic fields, ${\bfm E}$ and ${\bfm B}$ respectively, is determined by the Lorentz force
\begin{eqnarray}
\label{lorentz}
\frac{d{\bfm p}}{dt}=q({\bfm E} + {\bfm \beta}\times{\bfm B}),
\end{eqnarray}
where ${\bfm \beta}$ is the particle velocity normalized to the speed of light.
Any change in the energy of a 
particle originates from the work done by the ${\bfm E}$-field.
In most astrophysical environments, the local highly conducting plasma 
shorts out any large-scale electric field in the local fluid frame.
However, for particles which cross back and forth across
shocks there is no global frame in which 
all electric fields vanish. This is the underlying principle of diffusive-shock 
and shock-drift acceleration. In turbulent regions, it is the small-scale electric fields
associated with the plasma waves that accelerate particles.

It was shown by \citet{Tverskoi} that the evolution of the energetic
particle distribution can be described by the momentum diffusion equation 
\begin{equation}
\label{Tversk}
\pdiff{f(p,t)}{t} = \frac{1}{p^2}\pdiff{}{p}\left[
p^2 D(p)\pdiff{f(p,t)}{p}\right].
\end{equation}

\noindent which is valid for scattering times-scales much shorter than the crossing time of the physical system.

The form of the diffusion coefficient 
due to interactions with hydromagnetic waves
has been discussed in detail in numerous
publications \citep[e.g.][]{Kulsrud71,SchlickBook}.
We will focus on acceleration in a region of Alfv\'enic turbulence with
a one-dimensional power spectrum $W(k) \propto k^{-q}$, where 
$\delta B^2/8\pi = \int_{k_{\rm min}}^{k_{\rm max}}W(k) \diff k$,
with $k=2\pi/\lambda$. The wavenumbers $k_{\rm min}$ and $k_{\rm max}$ 
correspond to to the longest ($\lambda_{\rm max}$) and shortest 
($\lambda_{\rm min}$) wavelengths in the system respectively. The spectral index of the power spectrum is denoted by $q$. 

Neglecting numerical factors, and using the wave spectrum described above,
the momentum diffusion coefficient is~\citep{Schlickeiser89}
\begin{equation}
 D(p)\approx \beta_{\rm A}^2\frac{\delta B^2}{B_0^2} 
\left(\frac{r_{\rm g}}{\lambda_{\rm max}}\right)^{q-1}\frac{p^2c^{2}}{r_{\rm g}c}
\propto p^q.
\end{equation} 
where $r_g=pc/eB_0$ is the gyroradius of a particle, and $\beta_{\rm A}$ 
is the Alfv\'en velocity normalized to the speed of light.
This result is valid for particles with gyroradii
smaller than the correlation length of the field. The diffusion of particles with larger
gyroradii are essentially
independent of momentum since such particles interact with the entire spectrum
\citep{Tsytovich:72}. This results in an increase in acceleration times, 
a fact which we demonstrate numerically in
Section~\ref{numerics}. The associated mean free path, $L$, is 
\begin{equation}
 L\approx \frac{B_0^2}{\delta B^2}\left(\frac{r_g}{\lambda_{\rm max}}\right)^{1-q}r_g.
\end{equation}
Equation~\ref{Tversk} contains a systematic acceleration timescale given by 
\begin{equation}
\label{tacceq}
 t_{\rm acc} = \frac{p^2}{D(p)} 
\approx \beta_{\rm A}^{-2} \frac{L}{c}
\propto p^{2-q}.
\end{equation}

For the rest of this section we will consider exclusively 
a one-dimensional Kolmogorov spectrum of Alfv\'en waves
with $q=5/3$, and 
a turbulence level $\delta B^2/B_0^2=1.0$. While Kolmogorov turbulent
spectra are typical for such environments the actual value of the
turbulence level is more difficult to determine, particularly in radio lobes
where the magnetic Reynolds numbers are large. However, the results should be approximately accurate for gyroradii much less than the correlation length of the field since we calculate the Alfv\'en speed using the field's rms value.

For gyroradii larger than correlation length of the turbulence
the acceleration rate falls off as $p^{-2}$ \citep{Tsytovich:72,Ber90}\footnote{
Actually, the transition from $t_{acc}\propto p^{2-q}$ to 
$t_{acc}\propto p^{2}$ is smooth and starts to occur at gyroradii
less than the correlation length $r_g < \lambda_{\rm c}$, as shown in 
Section~\ref{numerics}.
Thus the results that follow can be considered as rough upper limits.}.
Hence, we consider the highest energy protons to be those 
which resonate with the largest wavelength in the system,
$r_g/\lambda_{\rm max}=1$. The largest wave is taken to be an order of magnitude less
than the size of the lobes $\lambda_{\rm max}\approx 0.1~R_{\rm lobe}$. 
We note that the gyroradius for a proton of energy $\varepsilon$ is
 $r_g\approx 10\left(\frac{\varepsilon}{10^{19}~{\rm eV}}\right) 
\left(\frac{B}{1~\mu {\rm G}}\right)^{-1}~{\rm kpc}$, 
such that for a radio lobe of size
$R_{\rm lobe}=100$~kpc, and with $B=1~\mu$G, 
protons of energy $10^{19}$~eV have gyroradii equal to $\lambda_{\rm max}$. 
These parameter are in a similar range to those expected in the lobes of Cen~A.
We also note that protons with energy $10^{20}$~eV (in the same 
magnetic field) have gyroradii comparable 
to the size of the system, making the giant lobes of
Cen~A marginal candidates for the acceleration of UHECR to these energies,
provided the particle has had sufficient time to reach these energies.
According to equation \ref{tacceq}, the corresponding acceleration 
time for such protons is 
$t_{\rm acc}(\varepsilon_{\rm max})\sim \lambda_{\rm max} /\beta_{\rm A}^2c$.

Using Faraday rotation measurements upper limits can be estimated for the density
in radio lobes. These are typically of the order $n_{\rm  p}=10^{-4}~{\rm cm}^{-3}$
\citep{BBR84}, although the proton density can in fact be much lower. 
Adopting this upper value for the density in the lobe,
the field and matter energy densities are roughly $0.025$~eV~cm$^{-3}$ and $5\times10^{4}$~eV~cm$^{-3}$ 
respectively, indicating an Alfv\'en speed of $\beta_{\rm A}=\left(U_{B}/U_{\rho}\right)^{1/2}=7\times 10^{-4}$ where $U_{B}=B^2/8\pi$
and $U_\rho=0.5n_{\rm p}m_{\rm p}c^2$.
Using these values, the acceleration time of a proton with energy $\varepsilon<\varepsilon_{\rm max }=10^{19}$~eV is 
\begin{equation}
\label{tacc}
t_{\rm acc} \approx 50\left(\frac{\beta_{\rm A}}{7\times 10^{-4}}\right)^{-2}\left(\frac{\lambda_{\rm max}}{10~{\rm kpc}}\right)^{2\over3}
\left(\frac{\varepsilon}{10^{19}~{\rm eV}}\right)^{1\over3}~{\rm Gyr},
\end{equation}
which is longer than the Hubble time. Clearly, if 
ultra-high energy protons are stochastically accelerated in radio lobes, the Alfv\'en speed must
be considerably larger, ie. lower density and/or higher magnetic fields. In fact,
as we will demonstrate, for Cen~A we require mildly relativistic Alfv\'en speeds
$\beta_{\rm A} \sim 0.1$.

We have so far neglected particle interactions with magnetosonic waves 
and acceleration due to transit-time damping (TTD).  It is well known that 
fast mode magnetosonic turbulence, when of comparable intensity 
to the Alfv\'{en} component, can present a more effective mechanism for particle
acceleration through TTD \citep[e.g][]{SchlickMill}, the acceleration 
time being shorter by a factor $\ln(\beta_{\rm A}^{-2})$. However, this is still 
not sufficient to explain the acceleration of UHECR using the above
canonical values for conditions in the lobes. Also, at Alfv\'en 
velocities $\beta_{\rm A} \gtrsim 0.05$ the acceleration times become comparable.

Furthermore, as has been pointed out by \cite{Eichler79} and \cite{achterberg81},
acceleration by fast-modes can become ineffective when the plasma beta
$\beta>m_{\rm e}/m_{\rm p}$, 
where $\beta = P_{\rm th}/U_{\rm B}$, with $P_{\rm th}$ the sum of partial pressures.
If the positron fraction of the plasma becomes significant, and the
corresponding proton density drops, the electrons will dominate the total pressure.
If $\beta$ satisfies the above constraint, 
most of the wave energy goes into near-thermal particles and 
the fast-modes are subject to rapid Landau damping. Assuming reasonable
parameters for radio lobes we find that $\beta\sim1$ and therefore fast modes may 
be effectively dissipated. However, this   
remains a topic of on-going investigation \citep[e.g.][]{YanLaz,petrosian06}. 

It may be the case that TTD makes a significant contribution to particle acceleration
in localized regions within the radio lobes. However, a large departure from
canonical parameters is required in order to accelerate particles to the highest 
energies and, as 
discussed above, it is likely that scattering from Alfv\'{e}nic turbulence is the 
dominant process for the highest energy particles. 
In the investigations presented here we have therefore focused exclusively on 
Alfv\'en waves and defer to further studies the inclusion of fast mode effects.

In Section~\ref{numerics}, we investigate the accuracy of the acceleration 
timescales for Alfv\'enic turbulence provided by \QLT~for both non-relativistic
and mildly relativistic Alfv\'en velocities, and for large 
$\delta B^2/B_0^2$. We emphasise here that the mildly relativistic models considered are highly speculative, involving ion densities three orders of magnitude below the upper limits derived from Faraday rotation measurements.

\section{Stochastic acceleration in Cen~A}
\label{CenA}

Cen~A is the nearest radio galaxy and one of the brightest extragalactic 
radio sources in the sky \cite[for a review see][and references therein]{Israel:1998}.
Concentrated within a circle of $20^{\circ}$ around the nucleus of Cen~A, 
are $10$ of the $27$ reported events above $10^{19.7}$~eV. While two appear
to come from the nucleus itself, many more coincide with its giant lobes
\citep{Mosk08}, which are approximately $\sim 600$~kpc in extent, and projected
on the sky have an angular size of $10^{\circ}$. These correlations may well
be deceptive with their sources possibly originating in the Centaurus supercluster
\citep{Gorb}, or simply scattered into our line of sight by the magnetic fields 
in the galaxy or lobes \citep{Lemoine}.

Nevertheless, being the closest radio-galaxy it has long been favoured as a 
source of UHECR \citep{Farrar:2000,Isola:2001}.
It was investigated in \citet{Hardcastle} whether these particles could be stochastically 
accelerated in the lifetime of the lobes. However, high Alfv\'en speeds were found to be necessary, in agreement with our discussion in Section~\ref{acceleration}. 
The actual value of the Alfv\'en velocity is uncertain.
Using minimum energy arguments, the magnetic fields are not expected to exceed 
$3\mu$G in the giant lobes of Cen~A \citep{Alvarez:2000}.

Due to uncertainty in the baryon fraction of the particle content
within the lobes, an accurate measurement of the density is a much more difficult task,
with little more than upper limit estimates possible.
While different upper limits have been given in different regions of the lobes
based on the soft X-ray thermal emission, there is majority agreement at $\sim10^{-4}$~cm$^{-3}$ 
\citep[for a review of different limits see sect. 7.3 of][]{Hardcastle}. 
However, these are only upper limits and the actual densities may be considerably
smaller.

Another uncertainty
lies in the correlation length of the turbulence. While larger values allow
for increased maximum proton energy, smaller values may favour rapid electron 
acceleration, provided the turbulent spectrum can extend to sufficiently
short wavelengths to resonate with the electrons.

The turbulence is generated on large scales by hydrodynamic instabilities.
Given that the projected physical 
extent of each lobe is approximately $250$~kpc $\times 100$~kpc, with
structure evident on all observed scales, a reasonable estimate of the correlation length is $\sim 10$~kpc.

The small-scale cutoff in the turbulent energy spectrum occurs at the scale $l_{\rm eq}$ at which equipartition between the kinetic and magnetic energies is reached. The correlation length may be in turn related to $l_{\rm eq}$ via the magnetic Reynolds number $R_{\rm m}$ according to $l_{\rm eq}\approx R_{\rm m}^{-1/2} l_{\rm turb}$~\citep{ruzmaikin89}. Assuming a characteristic bulk velocity of $0.1\,c$, and a temperature of $10^7$\,K, \cite{spitzer62} provides an expression for transverse resistivity which yields a magnetic Reynolds number $R_m\sim 10^{35}$. Hence, $l_{\rm eq}\ll r_{\rm g\, thermal}$, where $r_{\rm g\, thermal}$ is gyroradius of thermal protons.

With $\lambda_{\rm max}\approx10$~kpc and $B\approx3~\mu$G,
a maximum energy of $\varepsilon_{\rm max}\approx10^{19.7}$~eV is indicated for 
$r_g \lesim \lambda_{\rm max}$, which is in the energy range relevant for the
Auger events. The acceleration time for a particle of
energy $\varepsilon_{\rm max}$ is 
$t_{\rm acc}\approx 10^5~\beta_{\rm A}^{-2}~{\rm yrs}$,
where we have assumed a Kolmogorov spectrum $q=5/3$ and equipartition 
between the energy density in the turbulent and ordered magnetic field.

The age of the radio lobes in Cen~A is estimated to be $10^{7.5}$~yrs
\citep{Hardcastle}, which suggests 
Alfv\'en speeds $\beta_{\rm A} \gtrsim 0.1$ are necessary to accelerate 
particles to such high energies in the lifetime of the lobes. Again, assuming $B=3~\mu$G as
an upper limit indicates densities $\lesim 10^{-7.5}$~cm$^{-3}$
are necessary to achieve such values of $\beta_{\rm A}$.

\subsection{Electron acceleration}

As an interesting aside, one can calculate the
acceleration time of electrons in the lobes of Cen~A. We assume a turbulent spectrum with a uniform Kolmogorov power law extending to the length-scales required for resonance with radio emitting electrons.
However, we note that the uncertainty associated with such an assumption is significant.

The combined radiative cooling time for electrons via synchrotron 
in a $3~\mu$G magnetic field ($U_{\rm B}=0.2~{\rm eV~cm}^{-3}$) 
and via inverse Compton in the CMB radiation field ($U_{\rm CMB}\sim 0.3~{\rm eV~cm}^{-3}$) is 
\begin{eqnarray}
t_{\rm cool}&=&E_{e}/\frac{4}{3}c\sigma_{T}\beta_{e}^2 \gamma^{2}U_{\rm tot} \nonumber\\
&\approx& 0.83~\left(\frac{\varepsilon}{10^{12}~{\rm eV}}\right)^{-1}\left(\frac{U_{\rm tot}}{0.5~{\rm eV~cm}^{-3}}\right)^{-1}~{\rm Myr}
\end{eqnarray}
where $U_{\rm tot}=U_{\rm B}+U_{\rm CMB}$.
The corresponding acceleration time for electrons is
\begin{eqnarray}
\label{tacc_elec}
 t_{\rm acc} &\approx&  25\left(\frac{n_{\rm  p}}{10^{-4}~{\rm cm}^{-3}}\right)
\left(\frac{\lambda_{\rm max}}{10~{\rm kpc}}\right)^{{2\over3}}\nonumber\\
&\;&\;\;\; \times\;\left(\frac{U_{\rm B}}{0.2~{\rm eV~cm}^{-3}}\right)^{-{7\over6}}
\left(\frac{\varepsilon}{10^{12}~{\rm eV}}\right)^{1\over3} {\rm Myr}.
\end{eqnarray}
Thus cooling and acceleration timescales for an electron become 
comparable at a critical energy   
\begin{eqnarray}
\varepsilon_c&\approx&  
0.1~\left(\frac{n_{\rm  p}}{10^{-4}~{\rm cm}^{-3}}\right)^{-{3\over4}}
\left(\frac{\lambda_{\rm max}}{10~{\rm kpc}}\right)^{-{1\over2}}\nonumber\\
&\;&
\times\;
\left(\frac{U_{\rm B}}{0.2~{\rm eV~cm}^{-3}}\right)^{{7\over8}}
\left(\frac{U_{\rm tot}}{0.5~{\rm eV~cm}^{-3}}\right)^{-{3\over4}}{\rm TeV}
\label{EcritIC}
\end{eqnarray}
Below this energy, the acceleration or 
re-acceleration of electrons may not be safely neglected.

Recently, \citet{Hardcastle} have presented results of the radio spectra
from selected regions of Cen~A in the northern and southern lobes. 
These authors fit models to the spectra in the southern regions using 
an inverse Compton cooled power law spectrum,
where the spectral age increases further from the 
nucleus (both spectral ages being $\sim 10^{7.5}$~yrs).
In the northern regions the spectra obtained are different but 
can still be fit with a continuous
injection broken power law model. The northern
middle lobe is claimed to be responsible for the observed asymmetric
behaviour. The spectral breaks are found, in all regions, to occur 
in the range $10^9-10^{10}$~Hz (corresponding to photon energies of $10^{-6}-10^{-5}$~eV). 
In the $\sim\mu$G fields that are assumed in the lobes, this corresponds
to synchrotron radiating electrons with energies $\gtrsim 10^{10}$~eV,
which falls below the critical value $\varepsilon_c$, although could be consistent
with a lower magnetic field value.

Interestingly, our value for $\varepsilon_c$ was calculated using the upper 
limit for the number density. 
Lower densities such as those necessary for the stochastic acceleration of UHECR, 
increase the value of $\varepsilon_c$ by several orders of magnitude, suggesting
that stochastic acceleration may play an important role in the shape of
the electron spectrum at these energies. More self-consistent calculations,
similar to those of \citet{Kardashev} and \citet{Stawarz}, are warranted to 
investigate this further. 

Stochastic acceleration may also play an important role in the suggested 
re-acceleration of electrons observed near the hot-spots of other radio
galaxy jets \citep[e.g][]{Meisenheim}.

Finally, we reiterate that, since the acceleration
of $10^{10}$~eV electrons in a $3~\mu$G field
requires a Kolmogorov magnetic spectrum to span more than $9$ decades in 
length scale with a uniform power law, any softening of the turbulence
spectrum at large $k$ will reduce the effects of stochastic acceleration. 
Hence, the large Alfv\'en speeds required to accelerate UHECR, as
found in the previous section, are not necessarily
in conflict with the radio spectra.

\section{Numerical test-particle simulations}
\label{numerics}

The quasi-linear results of \cite{Schlickeiser89} used in the previous section 
are calculated assuming purely resonant interactions with a continuous
spectrum of non-relativistic Alfv\'en waves, $\beta_{\rm A} \ll 1$, 
with total energy density $U_{\rm turb} \ll U_{\rm B}$. 
Since our results have taken these energy densities to be
comparable, it is important to investigate the accuracy of the resulting expressions
for the acceleration time.

Numerical investigations of stochastic acceleration in Alfv\'enic turbulence have 
previously been performed by \citet{MichOst96} and \citet{MichOstSchlick99}. Their results show
good agreement with \QLT~although the simulations were limited in 
dynamical range, focusing on a single energy in a narrow range of
wavelengths. As yet, there has been no in depth numerical study of the
energy dependence of the acceleration times.

We calculate the acceleration timescales of test particles in low frequency Alfv\'enic
turbulence, using an \textit{a priori} field configuration which allows for
time-dependent waves. 

Two different approaches are adopted to determine these timescales. 

Direct simulation is carried out if acceleration is sufficiently rapid that the initial particle ensemble can be evolved from $\langle\gamma\rangle=10^5$ to $10^9$  within $4\times 10^7$ gyroperiods (measured at the initial energy). This is the case for the mildly relativistic turbulence models we consider in this work.

If the acceleration timescales are very long such that particle ensembles cannot be evolved over the range of interest in computationally viable simulations, we take \emph{snapshots} of mono-energetic particle populations over the range and rely on the slow dispersion of the injected particle delta function distribution to derive early-time diffusion coefficients for Gaussian fits to the particle distribution. This is the case for the non-relativistic models.

We remark here that our investigations are limited to protons within the sample range $\langle\gamma\rangle=10^5$ to $10^9$. Since we wish to evaluate the plausibility of Alfv\'{e}n mediated stochastic acceleration as a mechanism for driving protons to ultra-high energies, viable acceleration timescales over this range are a minimum necessary condition for acceleration from still lower energies. 

We note that protons with energies less than $100$~TeV will undergo stochastic
acceleration provided they have sufficient energy to satisfy the resonance condition 
with short-wavelength Alfv\'en waves. This condition is necessary to facilitate the pitch angle scattering required for maintenance of a quasi-isotropic distribution. 

We note that the accelerated particles require
some pre-acceleration process to lift them from the thermal pool. However, the consideration of pre-acceleration is beyond the
scope of this paper.

\subsection{Constructing the fields}

We have extended the test-particle transport code LYRA \citep{SOS09} 
to include time-dependent magnetic fields and corresponding time-dependent electric fields. 

A straightforward Lorentz transformation of magnetic 
fluctuations will, in general, lead to non-zero values of $\delta\mathbf{E}\cdot\mathbf{B}$. 
For the low frequency waves we consider, this is unphysical, and can lead to an 
unintended secular acceleration of 
the particle parallel to the local magnetic field \citep{Arzner}.

In this work, we shall consider two distinct field constructions which explicitly 
maintain $\delta\mathbf{E}\cdot\mathbf{B}=0$. The first is a one-dimensional 
field (in the sense that $\mb{k}$ is parallel or anti-parallel to $\mb{B}_0$). In the 
second construction, $\mb{k}$ is isotropically orientated in three-dimensions.

In each case, linear Alfv\'{e}n turbulence is imposed upon a mean field, $\mb{B}_0$, 
by superimposing a large number 
of linearly polarized modes, $\delta\mathbf{B}_j$ $(j=1..N)$, according to

\begin{equation}
\delta\mathbf{B}_j= A_j\widehat{(\mathbf{B}_0\times\mathbf{k}_j)}{\rm e}^{{\rm i}(\mathbf{k}_j\cdot\mathbf{r}-\omega_j t +\phi_j)} ,
\end{equation}

\noindent where $\hat{\mathbf{k}}_j$ and $\phi_j$ are chosen randomly within the 
dimensionality constraints imposed upon the construction of the field as described 
below. The turbulent component of the field then approaches homogeneity and  
isotropy in the limit of large $N$~\citep{batchelor}.
The random selection of Fourier phases, $\phi_j$, ensures the field has no 
large scale coherent structures, and
the amplitudes $A_j$ are chosen to create a Kolmogorov power spectrum. The
waves are taken to be purely Alfv\'enic such that 

\begin{equation}
\frac{\omega_j}{k_j c}=\beta_A |\cos\theta|
\end{equation}

\noindent with $\theta$ the angle between the wave-vector $\mathbf{k}_j$ 
and $\mathbf{B}_0$. Since the direction of $\mathbf{k}_j$ is determined
using two angles in spherical polar coordinates, chosen at random, 
the absolute value of the cosine must be taken to prevent a directional bias
in the fields.

\subsubsection{One-dimensional field}

In order to make contact with \QLT, we  consider a simple one-dimensional field 
construction similar to that used by~\cite{Kulsrud69}, \cite{Schlickeiser89} and others, whereby 
weak perturbations have wave-vectors purely parallel and anti-parallel to the mean 
field. Explicitly, each wavevector is determined by a random sign selection

\beq
\widehat{\mathbf{k}}^{\rm 1d}_j=\pm \widehat{\mathbf{B}}_0 .
\eeq

Vanishing $\delta\mathbf{E}\cdot\mathbf{B}$ is enforced by assuming a uniform linear 
polarization state for all modes. The polarization direction, 
$\widehat{\delta\mathbf{B}}_1$ say, is arbitrarily assigned perpendicularly to $B_0$ and  then fixed for all remaining modes according to

\beq
\widehat{\delta\mathbf{B}}_j = \widehat{\delta\mathbf{B}}_1 \quad \forall \quad j .
\eeq

\noindent Via a Lorentz transformation the associated electric field is

\begin{equation}
\delta\mathbf{E}_j= - \frac{\omega_j}{\mathbf{k}_j}A_j (\widehat{\mathbf{B}}_0\times\widehat{\delta\mathbf{B}_j}){\rm e}^{{\rm i}(\mathbf{k}_j\cdot\mathbf{r}-\omega_j t +\phi_j)} .
\end{equation}

\subsubsection{Three-dimensional field}

Numerical investigations on the effects of
anisotropy in the turbulent field have been reported by \cite{MichOstSchlick99}, by considering waves drawn randomly from within cones of differing opening angle about the mean field.
They find that the diffusion coefficients vary by less than an order of magnitude. 

However, it has also been argued by \cite{goldreich95, chandran00, cho02} that
Alfv\'enic turbulence is anisotropic, with eddies elongated along the (local) mean magnetic
field (ie. $k_\bot > k_\|$). \cite{YanLaz} have pointed out that this anisotropy
can result in inefficient gyroresonant acceleration
due to the reduction in the total power in parallel modes. The full importance of 
this effect is unclear in strongly turbulent fields, as may be prevalent in radio lobes.

For instance, there is observational evidence from polarization maps for chaotic fields in the southern lobe of Cen~A \citep{junkes93}. Furthermore, lobes such as
those of Cen~A  with morphologies characterized by extended filamentary structures~\citep{carilli89} are known to be weakly magnetized~\citep{clarke96} in the sense that $\delta B > B_0$. Consequentially, hydrodynamic turbulence, as might arise from high velocity injection of plasma into the lobes, will result in a strongly chaotic magnetic field component.

We now describe the construction of a magnetic field from modes with wavevectors 
directed isotropically in three-dimensions and a corresponding electric field 
such that \mbox{$\delta\mathbf{E}\cdot\mathbf{B}=0$}. 

Appealing to Ohm's law for an ideal plasma with infinite conductivity, we may write

\begin{equation}
\delta\mathbf{E}=-\delta\mathbf{u}\times\mathbf{B}, 
\label{eqn-ohm}
\end{equation}

\noindent where $\delta\mathbf{E}$, $\delta\mathbf{u}$, $\mathbf{B}$ are the total 
electric, bulk plasma velocity in units of $c$, and magnetic fields 
respectively in the observer frame.

Linearizing Faraday's law, and using the fact that linear Alfv\'en waves are 
non-compressional (ie. $\mathbf{k}_j\cdot\delta\mathbf{u}_j=0$), we have

\begin{equation}
\delta\mathbf{u}_j= -{\rm sgn}(\cos \theta)\frac{\beta_A}{B_0}\delta\mathbf{B}_j
\end{equation}

\noindent for the contribution to the velocity field due to each Alfv\'{e}n mode. Hence

\begin{equation}
\delta\mathbf{u}=-\frac{\beta_A}{B_0}(\mathbf{B}^+-\mathbf{B}^-),
\end{equation}

\noindent where $\mathbf{B}^+$($\mathbf{B}^-$) is the sum of all forward(backward) 
moving modes with respect to the mean field. From equation~\ref{eqn-ohm} then

\begin{equation}
\delta\mathbf{E}=-\delta\mathbf{u}\times\mathbf{B}_0 +2\frac{\beta_A}{B_0} (\mathbf{B}^+\times\mathbf{B}^-).
\label{eqn-e}
\end{equation}

\noindent Therefore, retaining the second order correction to $\delta\mathbf{E}$ 
given by the second term on the right-hand side of equation~\ref{eqn-e} provides a total electric 
field satisfying $\delta\mathbf{E}\cdot\mathbf{B}=0$. In 
the absence of this term, only $\delta\mathbf{E}\cdot\mathbf{B}_0=0$ and the electric field 
can still have a parallel component to the perturbing magnetic field, leading to 
rapid electrical acceleration of a charged particle propagating in the total B-field, $\mathbf{B}$.

\subsection{Equations of motion}

Given the time dependent fields $\mb{B}(\bfm{r})$ and $\delta\mb{E}(\bfm{r})$, the relativistic 
equations of motion for a particle of charge number $Z$, velocity $\bfm{\beta}\equiv \bfm{v}/c$, and rest mass $m$ are 

\beq
\frac{\mbox{d}\gamma \mb{\beta}}{\mbox{d}t}=\frac{Z e}{ m c} \left( \delta\mb{E}+\mb{\beta}\times \mb{B}\right) ,
\label{eqn-v1}
\eeq

\beq
\frac{\mbox{d}\mb{r}}{\mbox{d}t}=\mb{\beta}c ,
\label{eqn-v2}
\eeq

\n where $e$ is the electronic charge. In a magnetic field of mean magnitude $B_0$, the 
characteristic relativistic Larmor angular frequency is $\og\equiv Ze B_0/\gamma mc$ 
where $\gamma\equiv (1-\beta^2)^{-1/2}$. The 
corresponding maximal Larmor radius is $\rg\equiv\ogi \beta c$.

\subsection{Simulations}

To test the results presented in Section~\ref{acceleration}, 
we consider a homogeneous region of infinite extent 
with a mean field of $B_0=3\,\mu$G onto which is superimposed an isotropic  
turbulent component $\delta B$. The turbulence level of the field is determined via 
the parameter $(\sqrt{\langle\delta B^2\rangle}/B_0)^2\equiv(\delta B/B_0)^2$. 
The dynamic range of the turbulent field extends from 
$\lambda_{\rm min}=10^{-8}\,$kpc to $\lambda_{\rm max}=1\,$kpc 
and is resolved via 513 modes evenly 
spaced logarithmically. We have confirmed that our results are adequately converged at 
this mode density and dynamic range.

The Alfv\'en speed in the simulations 
is determined by the proton density according to 

\beq
\beta_{\rm A}=\frac{1}{c}\frac{B_0}{\sqrt{4\pi m_{\rm p} n_{\rm p}}}
\eeq

\n where $m_{\rm p}$ and $n_{\rm p}$ are the proton mass and number density respectively.

\begin{table}
\caption{\label{tab-params} Model parameters}
\begin{tabular}{ccccc}
Model & Dimension  &  $(\delta B/B_0)^2$ & $n_{\rm p}$ (cm${}^{-3}$) & $\beta_{\rm A}$  \\
\hline
       &    &      &       \\
\Aod   & 1  &  0.1 & $10^{-4}$ & 0.002 \\[0.2cm]
\Atd   & 3  &  0.1 & $10^{-4}$ & 0.002 \\[0.2cm]
\Bod   & 1  &  1.0 & $10^{-4}$ & 0.002 \\[0.2cm]
\Btd   & 3  &  1.0 & $10^{-4}$ & 0.002 \\[0.2cm]
\Cod   & 1  &  0.1 & $10^{-8}$ & 0.2   \\[0.2cm]
\Ctd   & 3  &  0.1 & $10^{-8}$ & 0.2   \\[0.2cm]
\Dod   & 1  &  1.0 & $10^{-8}$ & 0.2   \\[0.2cm]
\Dtd   & 3  &  1.0 & $10^{-8}$ & 0.2   \\[0.2cm]

\end{tabular}
\end{table}

We consider 8 different models in this work as defined by the parameters in 
Table~\ref{tab-params}. We consider non-relativistic turbulence $\beta_{\rm A}=0.002$ 
for both low and high turbulence levels $(\delta B/B_0)^2=0.1$ and $1.0$: 
models \Add and \Bdd respectively. We also consider mildly relativistic 
turbulence $\beta_{\rm A}=0.2$ for $(\delta B/B_0)^2=0.1$: model \Cdd; and 
for $(\delta B/B_0)^2=1.0$: model \Ddd. In each of these cases both one-dimensional 
and three-dimensional Alfv\'{en} turbulence is considered.

For the non-relativistic models \Add and \Bdd, the acceleration timescales are too long 
to capture the evolution of an injected particle distribution over a significant energy 
range with sufficient accuracy to resolve any statistical signal from numerical error. 
To see this, we have carried out benchmark integrations using a fifth order adaptive 
Runge-Kutta method with the accuracy tolerance set to $10^{-6}$. 

For model \Btd, it takes approximately one hour on an Intel~Xeon~2.66\,GHz~CPU to 
achieve an integration over $~6\times10^{-4}\,\tau_{\rm acc}(\gamma=10^9)$ for a proton 
at $\gamma=10^9$, or $~4\times10^{-5}\,\tau_{\rm acc}(\gamma=10^5)$ for a proton at $\gamma=10^5$.

Therefore, for the non-relativistic models \Add and \Bdd, we rely on snapshots of 
the acceleration rate at energies $\log_{10} \gamma=\{ 5,\, 6,\, 7,\, 8,\, 9\}$. 
Furthermore, we find that  integrations over 
$T=\{ 64,\, 32,\, 16,\, 8,\, 4\}\times 10^3 \omega_{0\,\rm g}^{-1}$ respectively, 
where $\omega_{0\,\rm g}(\gamma_0)$ is the relativistic Larmor frequency 
\emph{at the corresponding initial energy $\gamma_0$}, are sufficient for the 
momentum distributions to complete the initial ballistic transport phase and become 
diffusive. While $t\ll\tau_{\rm acc}$, the momentum distributions may be approximated via Gaussian profiles. This can be easily demonstrated for the special case of a flat wave spectrum ($q=1$) for which a solution to equation~\ref{Tversk} is provided in \cite{MichOst96}. We consider termination times for integration ranging from $4\times 10^{-4}\,\tau_{\rm acc}(\gamma=10^5)$ 
for particles injected at $\gamma=10^5$, to $1\times 10^{-2}\,\tau_{\rm acc}(\gamma=10^9)$ 
for particles injected at $\gamma=10^9$. 

\begin{figure}
  \centering
  \vbox{
    \psfrag{pp1}[Bl][Bl][1.0][+0]{\hspace*{-0.0cm}$\gamma$ }
    \psfrag{pp2}[Bl][Bl][1.0][+0]{\hspace*{-0.5cm}$\tau^{\rm err}_{\rm acc} (\lambda_{\rm c}/c)$ }    
    \psfrag{pp3}[Bl][Bl][0.8][+0]{\hspace*{-0cm}\Add}
    \psfrag{pp4}[Bl][Bl][0.8][+0]{\hspace*{-0cm}\Bdd}
    \psfrag{pp5}[Bl][Bl][0.8][+0]{\hspace*{-0cm}\Cdd}
    \psfrag{pp6}[Bl][Bl][0.8][+0]{\hspace*{-0cm}\Ddd}
    \psfrag{pp7}[Bl][Bl][0.8][+0]{\hspace*{-0.0cm}Sim}
    \psfrag{pp8}[Bl][Bl][0.8][+0]{\hspace*{-0.0cm}Fit}

    \includegraphics[width=160pt,angle=-90]{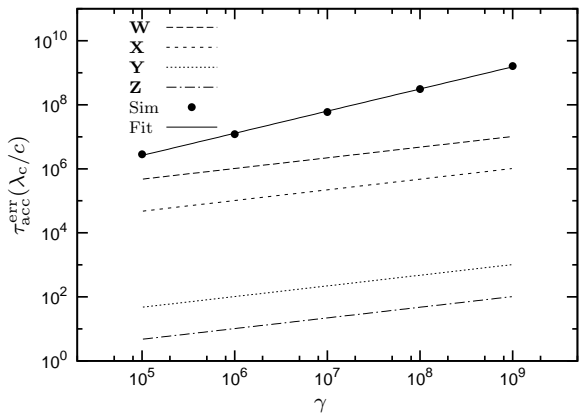}
  }
  \caption{Fictitious numerical acceleration $\tau^{\rm err}_{\rm acc}$ as a function of energy for three-dimensional turbulence with Runge-Kutta tolerance parameter $\epsilon=10^{-6}$. $\tau_{\rm acc}$ is derived from systematic drift of momentum distribution via $\langle\Delta\gamma\rangle/\Delta t $. Simulation results for 765 particle ensembles at 5 different initial energies are indicated by filled circles. A best fit power law, $908\gamma^{0.69}$ is shown as a solid line. The expected values from \QLT~for each of the case studies in this work are also represented via $C\gamma^{1/3}$: \Add - long dash - $C=10224$; \Bdd - short dash - $C=1022$; \Cdd - dot - 1.022; \Ddd - dot-dash - $C=0.1022$.}
\label{fig-3dzerova_acc}
\end{figure}

It has been confirmed that energy is conserved to high accuracy by the integrator by examining its 
behaviour over characteristic integration times in the special case $\beta_{\rm A}=0$ 
where $\tau_{\rm acc}\to \infty$. Under these conditions for three-dimensional turbulence 
with $(\delta B/B_0)=1.0$, energy is found to be conserved to within $~0.005\%$.
Figure~\ref{fig-3dzerova_acc} illustrates the systematic fictitious acceleration 
timescales experienced by particles in this instance. Reference lines indicating 
\QLT~predictions for the acceleration timescales for each of the cases considered here 
are also shown. It is evident that, assuming the numerical errors do not grow 
disproportionately for $\beta_{\rm A}>0$, sufficient numerical accuracy is employed 
to resolve the required signal from the statistical models.

For each of the \Add and \Bdd model snapshot energies, 1500 particles are
injected with random position and velocity orientation in a randomly generated field. 
No more than three particles are released in any given realization in an effort to 
reduce numerical noise. 

The mildly relativistic \Cdd and \Ddd models are less challenging numerically. This is because 
the acceleration timescales are much shorter and therefore the integration can be 
carried out over a relatively long time. We inject 500 particles at $\gamma=10^5$ 
for each of the one-dimensional and three-dimensional models and integrate 
continuously for $4\times 10^7\,\omega_{0\,\rm g}(\gamma=10^5)$. This is sufficient 
for acceleration to values of $\gamma$ in excess of $10^9$.

\begin{figure}
  \centering
  \vbox{
    \psfrag{ppx}[Bl][Bl][1.0][+0]{\hspace*{-0.75cm}$\Delta\gamma/\gamma_0$ }
    \psfrag{ppy}[Bl][Bl][1.0][+90]{\hspace*{-1.0cm}$f(\Delta \gamma/\gamma_0)$}
    \psfrag{pp3}[Bl][Bl][1.0][+0]{\hspace*{+0.0cm}Sim}
    \psfrag{pp4}[Bl][Bl][1.0][+0]{\hspace*{+0.0cm}Fit}

    \psfrag{pp1a}[Bl][Bl][1.0][+0]{\hspace*{+0.0cm}$\gamma_0=10^5$}
    \psfrag{pp2a}[Bl][Bl][1.0][+0]{\hspace*{+0.0cm}$\gamma_0=10^6$}
    \psfrag{pp3a}[Bl][Bl][1.0][+0]{\hspace*{+0.0cm}$\gamma_0=10^7$}
    \psfrag{pp4a}[Bl][Bl][1.0][+0]{\hspace*{+0.0cm}$\gamma_0=10^8$}
    \psfrag{pp5a}[Bl][Bl][1.0][+0]{\hspace*{+0.0cm}$\gamma_0=10^9$}

    \psfrag{pp1b}[Bl][Bl][1.0][+0]{\hspace*{+0.0cm}$\Delta t=17.66 (\lambda_c/c)$}
    \psfrag{pp2b}[Bl][Bl][1.0][+0]{\hspace*{+0.0cm}$\Delta t=88.28 (\lambda_c/c)$}
    \psfrag{pp3b}[Bl][Bl][1.0][+0]{\hspace*{+0.0cm}$\Delta t=441.4 (\lambda_c/c)$}
    \psfrag{pp4b}[Bl][Bl][1.0][+0]{\hspace*{+0.0cm}$\Delta t=2207 (\lambda_c/c)$}
    \psfrag{pp5b}[Bl][Bl][1.0][+0]{\hspace*{+0.0cm}$\Delta t=11035 (\lambda_c/c)$}

    \psfrag{pp1c}{}
    \psfrag{pp2c}{}
    \psfrag{pp3c}{}
    \psfrag{pp4c}{}
    \psfrag{pp5c}{}

    \psfrag{pp1d}{}
    \psfrag{pp2d}{}
    \psfrag{pp3d}{}
    \psfrag{pp4d}{}
    \psfrag{pp5d}{}

    \includegraphics[width=\columnwidth,angle=-0]{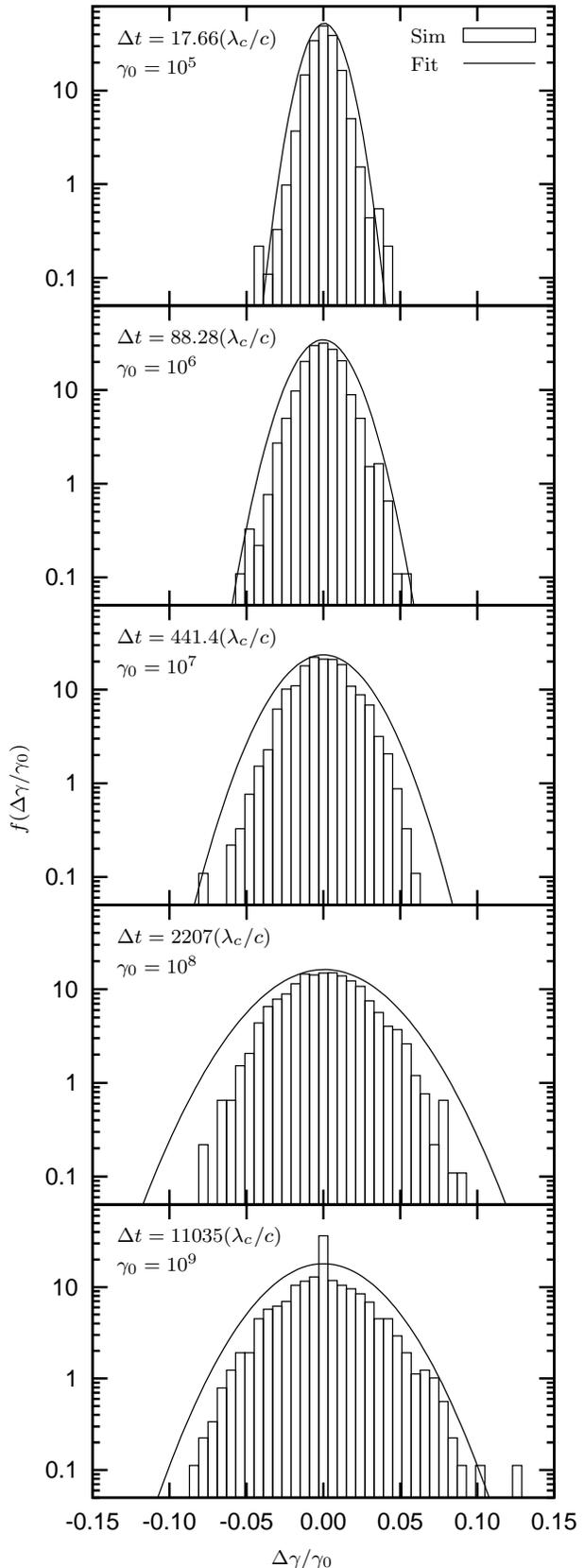}
  }
  \caption{Model \Aod: Histograms of relative energy change, $\Delta\gamma/\gamma_0$ (where  $\Delta\gamma\equiv\gamma-\gamma_0$), for ensembles of 1500 mono-energetic protons injected at $t=0$ with $\gamma=\gamma_0$. Solid line indicates a best-fit Gaussian distribution: the mean drift $\langle\Delta\gamma\rangle$ and variance $\langle(\Delta\gamma)^2\rangle$ are given in Table~\ref{tab-aod}.}
  \label{fig-1dlowturbhists}
\end{figure}

\subsection{Non-relativistic turbulence}

As previously remarked, for non-relativistic turbulence with $\beta_{\rm A}=0.002$, 
integration over a large energy range is not feasible computationally. Instead, 
ensembles of particles are injected at discrete energies over the range of 
interest and instantaneous acceleration times are inferred. This is achieved 
by allowing the injected delta function to relax to a Gaussian distribution and 
fitting its variance. 

In Figure~\ref{fig-1dlowturbhists}, snapshots of the particle distributions are 
illustrated for the \Aod model. The fitting parameters and errors are provided 
in Table~\ref{tab-aod} for reference. Evidently, while the mean drift upward in
energy is overwhelmed by the error, the standard deviation is well captured to 
within $\sim 5\%$. We have confirmed that the variance of the distribution 
does indeed vary linearly after a few scattering times, $\tau_{\rm s}$, ie. becomes diffusive. 
Furthermore, all simulated distributions achieve this diffusive phase 
of evolution with $\tau_{\rm s}<l_{\rm c}/c$ for $\rho<1$ and $\tau_{\rm s}\gesim l_{\rm c}/c$ for the highest energy test cases.

\begin{table}
\caption{\label{tab-aod} Model \Aod: Best-fit Gaussian distribution parameters.}
\begin{tabular}{ccccc}
$\gamma_0$ & \multicolumn{2}{c}{$\sqrt{\langle(\gamma-\langle\gamma\rangle)^2\rangle}/\gamma_0$}   & \multicolumn{2}{c}{$\langle\Delta\gamma\rangle/\gamma_0$} \\
\hline
&        &      &      &     \\
$10^5$ & 	 0.0107    &       $\pm$ 0.0005 & 	0.000506   &      $\pm$ 0.000583   \\[0.2cm]
$10^6$ & 	 0.0163    &       $\pm$ 0.0007 &      -0.000290   &      $\pm$ 0.000897   \\[0.2cm]
$10^7$ & 	 0.0239    &       $\pm$ 0.0011 & 	0.000031   &      $\pm$ 0.001297   \\[0.2cm]
$10^8$ & 	 0.0343    &       $\pm$ 0.0015 & 	0.000887   &      $\pm$ 0.001795   \\[0.2cm]
$10^9$ & 	 0.0314    &       $\pm$ 0.0025 & 	0.000065   &      $\pm$ 0.003111   \\[0.2cm]
\end{tabular}
\end{table}

The momentum diffusion coefficient is approximated by

\beq 
D_p = \frac{\langle\Delta \gamma^2\rangle}{2\Delta t}
\eeq

\n where we have used the fact that the mean drift in the distribution is 
negligible over the integrated times ie. 
$\Delta \gamma=\gamma-\langle\gamma\rangle\approx\gamma-\gamma_0$. 
From this expression, the acceleration time, $\tau_{\rm acc}$, is then obtained using

\beq
\tau_{\rm acc}=\frac{\gamma_0^2}{D_p} .
\eeq

\begin{figure}
  \centering
  \vbox{
    \psfrag{pp1}[Bl][Bl][1.0][+0]{\hspace*{-0.0cm}$\langle\gamma\rangle$}
    \psfrag{pp2}[Bl][Bl][1.0][+0]{\hspace*{-0.5cm}$\tau_{\rm acc} (\lambda_{\rm c}/c)$ }    

    \psfrag{pp3}[Bl][Bl][1.0][+0]{\hspace*{-0.7cm}QLT}
    \psfrag{pp4}[Bl][Bl][1.0][+0]{\hspace*{-0.7cm}\Aod Sim}
    \psfrag{pp6}[Bl][Bl][1.0][+0]{\hspace*{-0.7cm}\Aod Fit}
    \psfrag{pp5}[Bl][Bl][1.0][+0]{\hspace*{-0.7cm}\Atd Sim}
    \psfrag{pp7}[Bl][Bl][1.0][+0]{\hspace*{-0.7cm}\Atd Fit}

    \includegraphics[width=160pt,angle=-90]{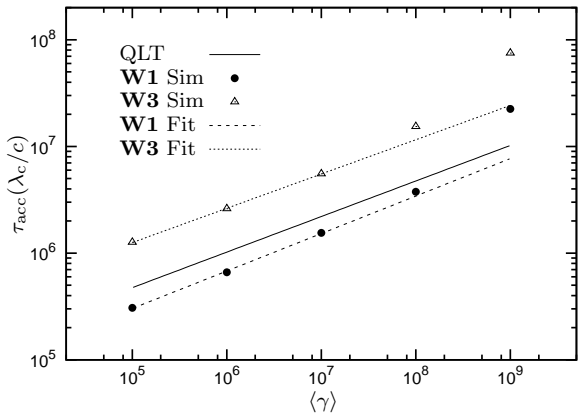}
  }
  \caption{Models \Add: Acceleration $\tau_{\rm acc}$ as a function of energy. $\tau_{\rm acc}$ is derived from best fits of Gaussian profiles to the standard deviation of the momentum distributions via $\gamma_0^2/D_p$ where $D_p\equiv\langle\Delta \gamma^2\rangle/2\Delta t$. The expected value from \QLT, $10224\gamma^{1/3}$, is represented by the solid line. Note that for $\rho \equiv r_{\rm g}/\lambda_{\rm c} {\protect \gesim} 1$ the efficacy of the acceleration process is damped. The best fit to the data for $\rho < 1$ ($\langle\gamma\rangle<3.6\times 10^8$) is given by $4531\gamma^{0.36}$ (dashed line) for \Aod, and $19138\gamma^{0.36}$ (dotted line) for \Atd.}
  \label{fig-lowturbdiff}
\end{figure}

 The results for the weakly turbulent models \Aod and \Atd are plotted in 
Figure~\ref{fig-lowturbdiff}. In both cases an index of $0.36$ is found 
for $\tau_{\rm acc}$ using a best fit containing all points excluding that
for the $10^9$ ensemble.
This is in close agreement with the theoretical value of $1/3$ 
dictated by \QLT, although the index does appear to approach this value
asymptotically for lower rigidities.  
Notably the three-dimensional field presents 
acceleration times that are a factor of $~4$~larger than the one-dimensional 
field case. This difference occurs due to the $\cos\theta$ dependence of the
electric field strength for oblique Alfv\'en waves.
The fact that the same normalization
for the total turbulent power was used for both simulations,
$\delta B^2/8\pi$ naturally accounts for the longer acceleration times.
Comparing the results from the one-dimensional simulations with the \QLT~result, 
we note that the numerically determined acceleration time is faster by a 
factor of almost $2$. 
This may be due to the fact that the particles in our simulations can
resonate with higher harmonics not accounted for in \QLT, or simply that
the turbulence level is not low enough. Nevertheless, the results in all
cases are in good agreement to order of magnitude.

It is also observed that for $\rho\gesim 1$ ($\gamma\gesim 3.6\times 10^8$) 
the acceleration time strongly diverges from the theoretical result as 
particles' gyromotions exceed the scale at which resonant waves are present in the field.

Regarding the acceleration of UHECR, this emphasizes the fact that
the maximum energy is quite typically governed by the correlation length of
the field.

\begin{figure}
  \centering
  \vbox{
    \psfrag{pp1}[Bl][Bl][1.0][+0]{\hspace*{-0.0cm}$\langle\gamma\rangle$}
    \psfrag{pp2}[Bl][Bl][1.0][+0]{\hspace*{-0.5cm}$\tau_{\rm acc} (\lambda_{\rm c}/c)$ }    

    \psfrag{pp3}[Bl][Bl][1.0][+0]{\hspace*{-0.7cm}QLT}
    \psfrag{pp4}[Bl][Bl][1.0][+0]{\hspace*{-0.7cm}\Bod Sim}
    \psfrag{pp6}[Bl][Bl][1.0][+0]{\hspace*{-0.7cm}\Bod Fit}
    \psfrag{pp5}[Bl][Bl][1.0][+0]{\hspace*{-0.7cm}\Btd Sim}
    \psfrag{pp7}[Bl][Bl][1.0][+0]{\hspace*{-0.7cm}\Btd Fit}

    \includegraphics[width=160pt,angle=-90]{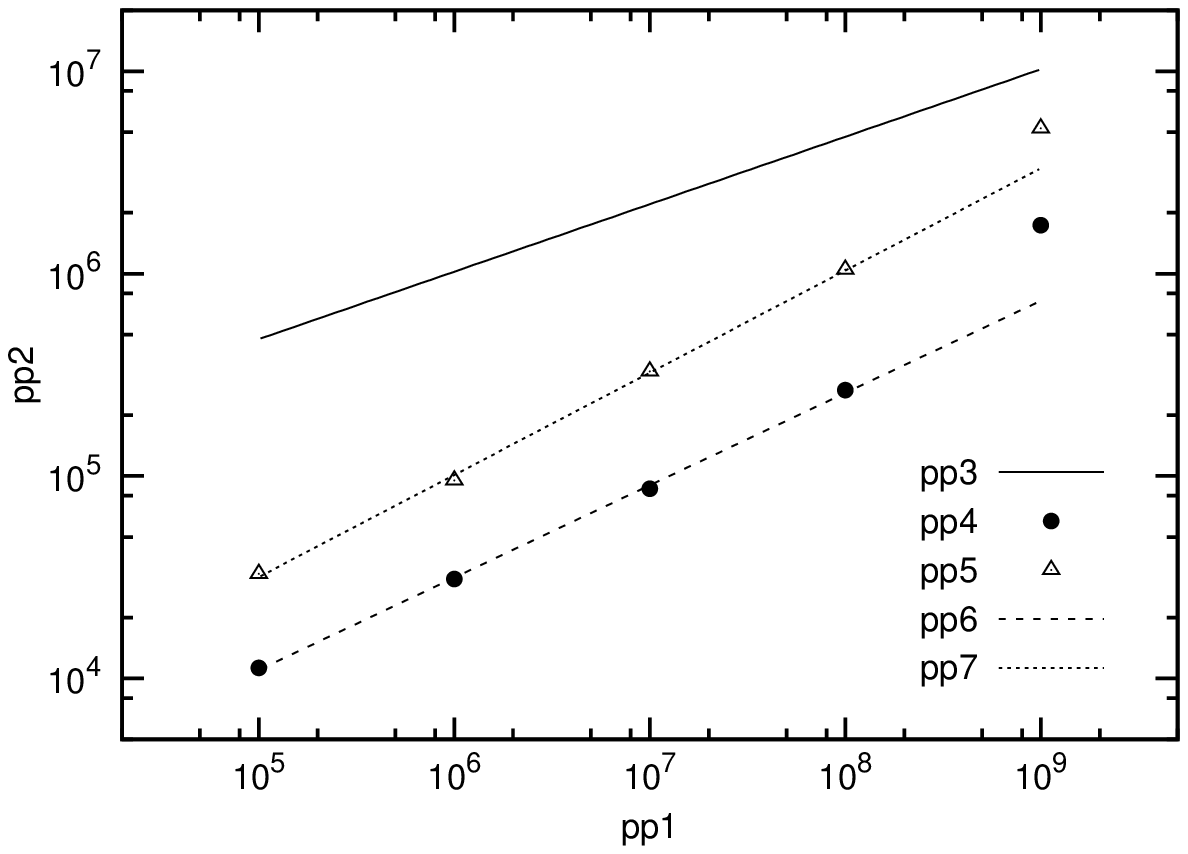}
  }
  \caption{Models \Bdd: Acceleration $\tau_{\rm acc}$ as a function of energy. $\tau_{\rm acc}$ is derived from best fits of Gaussian profiles to the standard deviation of the momentum distributions via $\gamma_0^2/D_p$ where $D_p\equiv\langle\Delta \gamma^2\rangle/2\Delta t$. The expected value from \QLT, $1022\gamma^{1/3}$, is represented by the solid line. Note that for $\rho \equiv r_{\rm g}/\lambda_{\rm c} {\protect \gesim} 1$ the efficacy of the acceleration process is damped. The best fit to the data for $\rho < 1$ ($\langle\gamma\rangle<3.6\times 10^8$) is given by $58\gamma^{0.46}$ (dashed line) for \Bod, and $95\gamma^{0.50}$ (dotted line) for \Btd.}
  \label{fig-highturbdiff}
\end{figure}

For models \Bod and \Btd, where the turbulence level is high, the agreement with \QLT~is
less compelling, as illustrated in Figure~\ref{fig-highturbdiff}. Here, the indices 
found are 0.46 and 0.50 for the one- and three-dimensional cases respectively, 
and the experimentally determined acceleration times in both cases 
are also an order of magnitude faster than predicted in \QLT. The 
divergence from \QLT~values with increasing turbulence level is not surprising, 
as formally \QLT~is valid in the limit of weak perturbations to the mean 
field. Again, the three-dimensional field shows acceleration times longer than 
the one-dimensional case by factor of $\sim 4$.

\subsection{Mildly relativistic turbulence}

\begin{figure}
  \centering
  \vbox{
    \psfrag{pp1}[Bl][Bl][1.0][+0]{\hspace*{-0.25cm}$t (\lambda_{\rm c}/c)$ }
    \psfrag{pp2}[Bl][Bl][1.0][+0]{\hspace*{-0cm}$\langle\gamma\rangle$ }    
    \psfrag{pp3}[Bl][Bl][1.0][+0.0]{\hspace*{-0.7cm}QLT}
     \psfrag{pp4}[Bl][Bl][1.0][+0.0]{\hspace*{-0.7cm}\Cod}
     \psfrag{pp5}[Bl][Bl][1.0][+0.0]{\hspace*{-0.7cm}\Ctd}
     \psfrag{pp6}[Bl][Bl][1.0][+0.0]{\hspace*{-0.7cm}$\rho=1$}

    \includegraphics[width=160pt,angle=-90]{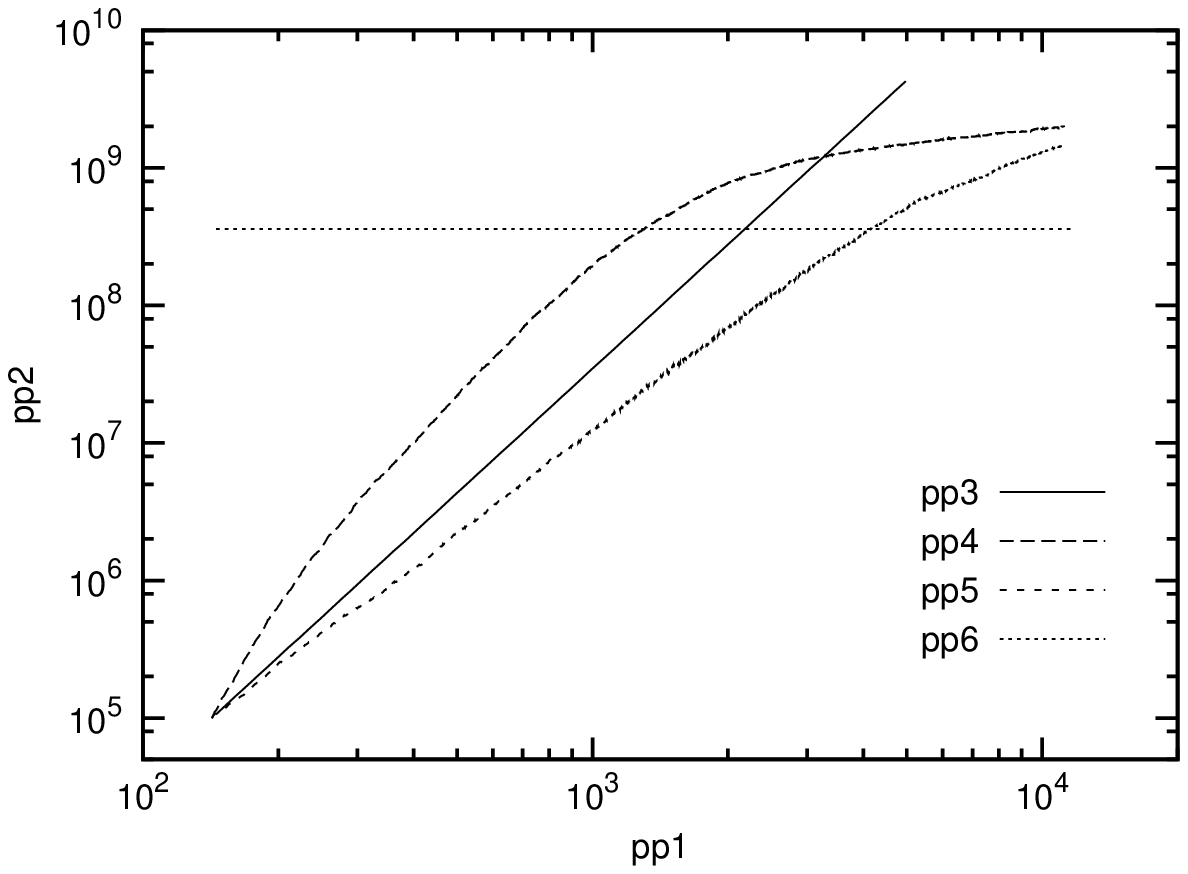}
  }
  \caption{Models \Cdd: Mean energy against time for ensembles of 510 particles injected at $\gamma_0=10^5$. Reference line indicates $\rho=1$.\QLT~result is $\langle\gamma\rangle=0.03465 t^3$. \Cod numerical results are indicated by long-dashed line;  \Ctd numerical results are indicated by short-dashed line. The critical energy $3.6\times 10^8$ at which $\rho=1$ is indicated by a dotted line.}
  \label{fig-lowturbhighva}
\end{figure}

For the mildly relativistic models \Cdd and \Ddd, with $\beta_{\rm A}=0.2$, $n_{\rm p}$ has been set to $10^{-8}$\,cm$^{-3}$, a value four orders of magnitude below 
Faraday rotation upper limits. While this value may be extreme, we wish to demonstrate the conditions under which second-order Fermi acceleration via pure Alfv\'{e}nic turbulence is viable in our idealized model. We note that such low densities
require a large fraction of positrons in order to safely neglect the influence of fast modes, as discussed in Section~\ref{acceleration}.

We find that for models \Cdd and \Ddd, the systematic acceleration is strong enough that we may simulate the acceleration 
of a single ensemble of particles from a low injection energy, $\gamma_0=10^5$, 
up to values of $\lesim 10^{10}$. Furthermore, fewer particles are required 
in order to derive a smooth signal. In practice this corresponds to 
integrations of 500 particles per model over $4\times 10^7\,\omega_{\rm g}^{-1}(\gamma_0)$.

\begin{figure}
  \centering
  \vbox{
    \psfrag{pp1}[Bl][Bl][1.0][+0]{\hspace*{-0.25cm}$t (\lambda_{\rm c}/c)$ }
    \psfrag{pp2}[Bl][Bl][1.0][+0]{\hspace*{-0cm}$\langle\gamma\rangle$ }    
    \psfrag{pp3}[Bl][Bl][1.0][+0.0]{\hspace*{-0.7cm}QLT}
     \psfrag{pp4}[Bl][Bl][1.0][+0.0]{\hspace*{-0.7cm}\Dod}
     \psfrag{pp5}[Bl][Bl][1.0][+0.0]{\hspace*{-0.7cm}\Dtd}
     \psfrag{pp6}[Bl][Bl][1.0][+0.0]{\hspace*{-0.7cm}$\rho=1$}

    \includegraphics[width=160pt,angle=-90]{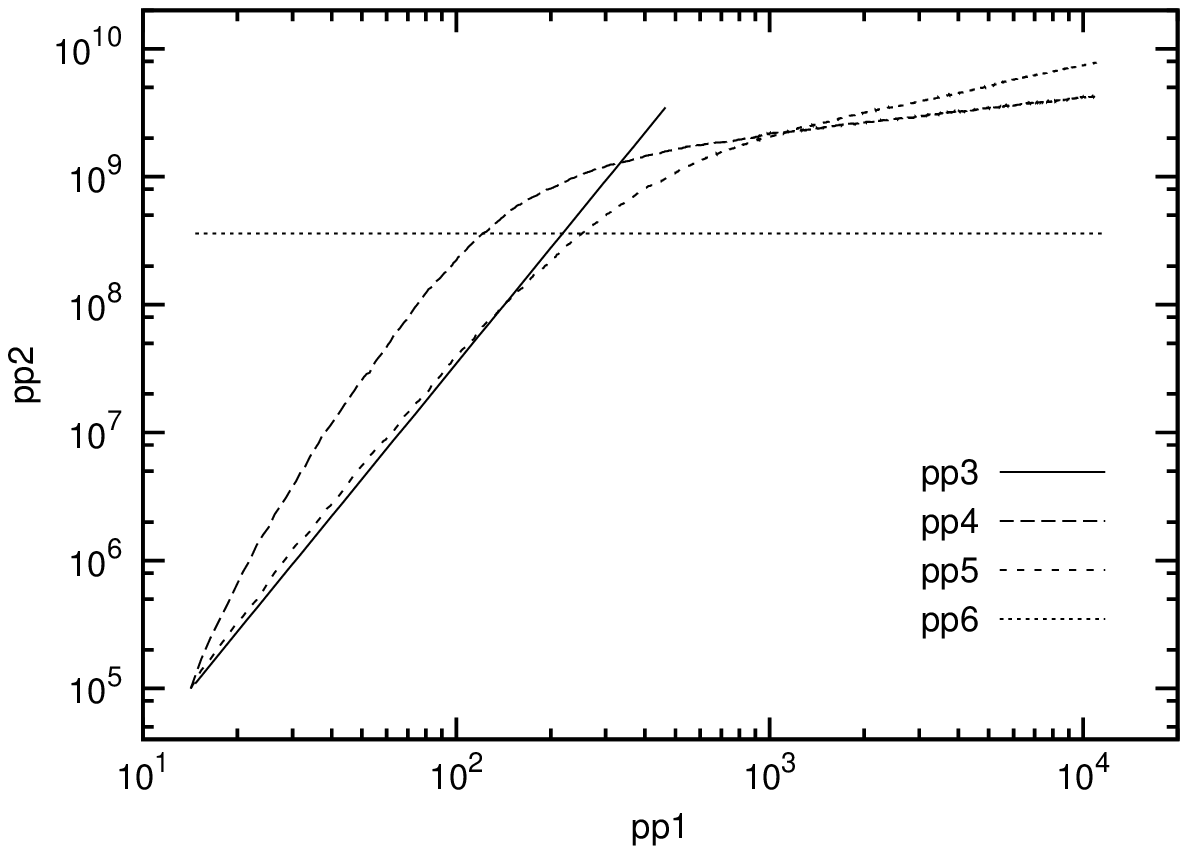}
  }
  \caption{Models \Ddd: Mean energy against time for ensembles of 500 particles injected at $\gamma_0=10^5$. Reference line indicates $\rho=1$.\QLT~result is $\langle\gamma\rangle=34.65 t^3$. \Dod numerical results are indicated by long-dashed line;  \Dtd numerical results are indicated by short-dashed line. The critical energy $3.6\times 10^8$ at which $\rho=1$ is indicated by a dotted line.}
  \label{fig-highturbhighva}
\end{figure}

In these cases, the one dimensional field models, \Cod and \Dod, exhibit more rapid acceleration than predicted by \QLT, with the low and high turbulence cases showing similarly curved profiles above the predicted $\langle\gamma\rangle \sim t^3$ power law.
However, due to the relatively rapid acceleration in the plane
perpendicular to the mean field, isotropy of the particle distribution cannot be
maintained. As such, comparisons with the results of \QLT~are misleading.

The mean energy plots for the three-dimensional field models, \Ctd and \Dtd, have flatter profiles. The high turbulence model is in excellent agreement with \QLT~and the lower turbulence level case has a somewhat slower growth rate than predicted by theory.

As before, in all cases, once $\rho\gesim 1$, the acceleration is severely damped by lack of resonant scales in the field.

\section{Conclusions}
\label{conclusion}

The stochastic acceleration of UHECR due to Alfv\'{e}nic turbulence in the lobes of giant radio galaxies has been 
considered. The work is motivated by the recently reported correlation between a 
number of the highest energy events observed by the Pierre Auger observatory 
and nearby AGN. Making simple estimates,
based on the results of \QLT, we have investigated the maximum energies
that can be achieved with the use of reasonable parameters in radio lobes.

While the maximum possible energy that can be achieved in each lobe is
primarily determined by the coherence length of the magnetic field
in the lobes, the timescale to achieve this is largely determined by the Alfv\'en speed. 
Assuming homogeneity of all quantities 
throughout the entire lobe, we have placed constraints on the value
of the Alfv\'en speed necessary to stochastically accelerate particles 
$10^{18}$~eV within the lifetime of the lobes.

We have demonstrated that Alfv\'en speeds in excess of $0.1~c$ are 
required in Centaurus~A to accelerate cosmic rays above $10^{19}$~eV, although
heavier nuclei could, in principle, achieve even higher energies.
Also, should the composition of UHECRs with energies above
$10^{19}$~eV be dominated by heavier nucleii, the observed UHECRs could
be accelerated in the lobes of Cen~A with an Alfv\'en speed an order of magnitude 
smaller.
While the absolute value of the magnetic field strength in the lobes is
quite well constrained by minimum energy arguments to
fall in the range $\sim 0.1-3~\mu$G, 
the densities referred to in the literature are
merely upper limits, and indeed the proton number density could be considerably
lower. 

Provided the turbulence spectrum steepens considerably at 
smaller wavelengths, such high Alfv\'en velocities are not in conflict
with the synchrotron radio emission from the non-thermal electrons.
Conversely, should the Kolmogorov spectrum extend to sufficiently small 
wavelengths, the resulting rapid acceleration time of electrons might
present a problem to the standard picture of an injected power law 
with a cooling break.

We have not allowed for transit-time damping (TTD) in the studies presented here since fast magnetosonic waves are not included in the turbulent field. While this simplifies the model and its interpretation, it may be justified by considering the efficacy of energy losses from these modes through heating of near-thermal electrons via Landau resonance effects. A comprehensive study of the linear theory of stochastic acceleration using 
a fully relativistic approach is, to our knowledge, still lacking, but a naive extension of the non-relativistic results
to mildly relativistic phase velocities suggest that acceleration by Alfv\'en waves
dominates above $\beta_{\rm A} \sim 0.1$.

The numerical results presented in this work demonstrate purely stochastic acceleration of relativistic charged particles through interaction with linearly polarized Alfv\'{e}n waves. The ideal MHD constraint, $\delta\mathbf{E}\cdot\mathbf{B}=0$, is trivially maintained for fields with perturbations directed parallel to the mean field (1D) and by deriving an appropriate electric field from inferred velocity perturbations via Ohm's law for an ideal plasma (3D). For both these one-dimensional and three-dimensional field configurations we consider low and high turbulence levels, $(\delta B/B_0)^2=0.1,\, 1.0$, for non-relativistic and mildly relativistic turbulence, $\beta_{\rm A}=0.002,\,0.2$.

For non-relativistic turbulence, we find that agreement with \QLT~is good at low turbulence levels. The three-dimensional acceleration timescales are found to be in excess of \QLT~by a factor of $\sim 2$. At higher turbulence levels, this agreement is less pronounced with the power law dependence of the acceleration time on the energy becoming steeper. Over the range considered in this work, the absolute acceleration times are also shorter than predicted by \QLT~with approximately the same relative difference between the one-dimensional and three-dimensional cases. 

Notably, above the critical energy corresponding to
resonance with the largest wave in the system, $\rho=1$, the particles interact with the the entire Alfv\'{e}n spectrum such that the diffusion becomes almost independent of energy. This dramatically reduces the efficacy of the acceleration process.

For the simulations with mildly relativistic Alfv\'en speeds, 
the mean energy of the particle ensembles shows reasonable agreement with \QLT~up to $\rho=1$ for both low and high turbulence levels. The one-dimensional cases have more rapid acceleration than predicted by theory, but the distributions become anisotropic,
and comparisons with \QLT~are not applicable.

Given that the results of the numerical simulations are in reasonable agreement
with the results of \QLT, in the mildly relativistic regime, the conditions necessary
for stochastic acceleration of UHECR in Cen~A can be satisfied, provided the 
correlation length of the magnetic field is sufficiently large, and that a particle
can remain inside the lobe for a sufficiently long length of time. Paramount to 
this result, is the relative low density of baryons in the lobes, which must fall
considerably below the upper limits inferred from observations.

\section*{Acknowledgments}

The authors would like to thank J.G. Kirk, P. Duffy and F. Rieger for their helpful
advice and comments. SOS is grateful to the SFI Research Frontiers Programme (RFP) for supporting this research. BR gratefully acknowledges support from the Alexander 
von Humboldt foundation.

The authors wish to acknowledge the SFI/HEA Irish Centre for High-End Computing (ICHEC) and the Centre for Scientific Computing \& Complex Systems Modelling (SCI-SYM) at DCU for the provision of computational facilities and support.

\label{lastpage}

\end{document}